\documentclass[submission,copyright,creativecommons]{eptcs}
\providecommand{\event}{DCM 2015}

\usepackage{amsmath}
\usepackage{amsfonts}
\usepackage{amssymb}
\DeclareMathOperator\artanh{artanh}
\DeclareMathOperator\arsinh{arsinh}

\title{Free fall and cellular automata}

\author{Pablo Arrighi
\institute{Aix-Marseille Univ., LIF, F-13288 Marseille Cedex 9, France.}
\email{pablo.arrighi@univ-amu.fr}
\and
Gilles Dowek
\institute{Inria, 
LSV, ENS-Cachan, 61 Avenue du Pr\'esident Wilson, 94230 Cachan, France.}
\email{gilles.dowek@inria.fr}}
\def\titlerunning{Free fall and cellular automata}
\def\authorrunning{P. Arrighi \& G. Dowek}

\begin{document}
\maketitle

\begin{abstract}
  Three reasonable hypotheses lead to the thesis that physical
  phenomena can be described and simulated with cellular automata.  In
  this work, we attempt to describe the motion of a particle upon
  which a constant force is applied, with a cellular automaton, in
  Newtonian physics, in Special Relativity, and in General
  Relativity. The results are very different for these three theories.
\end{abstract}

\section{Introduction}

Three reasonable hypotheses---homogeneity in time and space, bounded
velocity of propagation of information, and bounded density of
information---lead to the thesis that physical phenomena can be
described and simulated with cellular automata. This implication has
in fact been formalized into a theorem both in the classical
\cite{Gandy} and the quantum case \cite{ArrighiGANDY}, albeit in flat
space.

Further evaluating this thesis leads to the project of selecting
specific physical phenomena and attempting to describe them with
cellular automata. In this work, we consider a particle upon which a constant force is applied---as induced by the first order approximation of a gravitational field. We do so in three different settings: Newtonian physics, Special Relativity, and General Relativity. We seek to capture each of these motions as a Cellular Automaton. The results are very different for these three theories.

\section{Motion in cellular automata}

Recall that the configurations of a 1D cellular automaton are
functions from $\mathbb Z$ to a finite set of states $\Sigma$, which
includes a distinguished quiescent state $q$.  The evolution of the
cellular automaton is a function $F$ mapping configurations to
configurations. It has to be causal and homogeneous, that is there must
exist a radius $r$ and a local function $f$ such that for all $i$,
$(F(\delta))_i = f(\delta_{i-r}, ...,, \delta_{i-1}, \delta_i,
\delta_{i+1}, ..., \delta_{i+r})$.

Consider a temporal step $\varepsilon$, a spatial step $\Delta$, and a
discrete motion $\tilde{y}$, that is a function from $\varepsilon
{\mathbb N}$ to $\Delta {\mathbb Z}$.  A configuration $\delta$ is
said to represent a particle at position $k \Delta$ if $\delta_k \neq
q$ and for all $i \neq k$, $\delta_i = q$.  A transition function $F$
represents a discrete motion $\tilde{y}$ if there exists an initial
configuration $\delta$ such that for all $k$, $F^k(\delta)$ represents
a particle at $\tilde{y}(k\varepsilon)$. A standard reference on
cellular automata for constructing signals approximating different
functions is \cite{Mazoyer199953}.

In each of the following three sections we proceed by first
calculating the continuous motion $y(t)$, that is the position as a
function of time. We then construct the cellular automata for
$\tilde{y}$, if it exists. The differences between the three cases are
highlighted.

\section{Free fall in Newtonian physics}

We begin with the standard case of Newtonian physics. The choice of units and notations will carry through in the next sections. In Newtonian physics, the gravitational force applied by a body of
mass $M$ upon a particle of mass $M'$ at a distance $d$ is
$$F = {\cal G} \frac{MM'}{d^2}$$ 
Multiplying both sides by ${\cal G}/c^4$, where $c$ is the speed of
light, introducing notations $m = ({\cal G}/c^2) M$ which is the mass
of the body expressed in meters, $m' = ({\cal G}/c^2) M'$ which is the
mass of the particle expressed in meters, and $f = ({\cal G}/c^4) F$
which is the force expressed as a scalar without dimension, we get
$$f = m' \frac{m}{d^2}$$
Consider a particle whose initial distance to the body is $R$ and 
initial velocity is zero and let $y$ be such that $d = R - y$,
we have
\begin{align}
f = m' \frac{m}{(R - y)^2} \label{eq:NP}
\end{align}
To define free fall, we approximate this force by 
\begin{align}
f = m' \frac{m}{R^2}\label{eq:NPapprox}
\end{align}
that is, introducing the notation $g = m/R^2$
$$f = m' g$$
For example, the mass of the Earth is $M = 5.97~10^{24}~\mbox{kg}$,
so $m = ({\cal G}/c^2) M = 4.42~10^{-3}~\mbox{m}$.  The radius of the
Earth is $R = 6.37~10^6~\mbox{m}$, so $g = m/R^2 =
1.09~10^{-16}~\mbox{m}^{-1}$. Note that $g c^2 = 9.81~\mbox{m}
\mbox{s}^{-2}$ as expected.

When such a constant force is acting on a particle of mass $M'$, its
acceleration $A$ is given the equation
$$M' A = F$$
Multiplying both sides with ${\cal G}/c^4$ and introducing 
the notation $a = A/c^2$, which is the acceleration of the particle 
expressed in m$^{-1}$, we get 
$$m' a = f = m' g$$ 
thus 
\begin{align}
a = g \label{eq:NP1}
\end{align}
from which we get
$$v = g t$$
where $v = V/c$ is the velocity expressed as
a scalar with no dimension and $t = cT$ is time expressed in meters, 
and 
$$y= \frac{1}{2} gt^2$$
Thus the spacetime trajectory of this particle is a parabola. 

It is easy to prove that no cellular automaton can simulate such a
motion: as the velocity of the particle increases linearly with time,
the difference between $y$ at some time step and at the next time step
increases linearly with time. Thus, the evolution is not local.
Moreover, to be able to compute $y$ at the next time step from $y$ at
some time step, we need to know the velocity of the particle and it is
then natural to express this velocity as part of the state of the
cell. But then, as velocity is not bounded, the state space cannot be
kept finite, even if velocity is defined with a finite precision.

\section{Constant force in Special Relativity}

In Special Relativity, neither of these problems occurs: velocity is
bounded, hence the evolution is local.  And if the velocity is known
with a finite precision, a finite state space suffices.  Still,
another worry remains. If the velocity at some time step is computed
from the velocity at the previous one, and both velocities are
approximate, errors can accumulate. As we shall see, it is possible to
circumvent this problem, and have a non divergent discretization of
the trajectory of the particle.

In Special Relativity, the proper acceleration \cite{WheelerSpacetime} 
of a particle is
$$A = \frac{1}{\sqrt{1-V^2/c^2}^3} \frac{dV}{dT}$$
dividing both sides by $c^2$, we get 
$$a = \frac{1}{\sqrt{1-v^2}^3} \frac{dv}{dt}$$
We assume that the force is as in Newtonian physics: $m' a = f = m' g$, so $a = g$, that is
$$\frac{1}{\sqrt{1-v^2}^3} \frac{dv}{dt} = g$$
This assumption, however, is now better understood as ``constant force'' than ``free fall''. We get 
\begin{align}
\frac{dv}{dt} = g \sqrt{1-v^2}^3.\label{eq:accel}
\end{align}
Solving this equation, we get 
\begin{align}
v = \frac{g t}{\sqrt{1 + (gt)^2}}\label{eq:speed}
\end{align}
as the reader may check by differentiating \eqref{eq:speed} and comparing the result with \eqref{eq:accel} with $v$ substituted by \eqref{eq:speed}.
Then
$$y = \frac{1}{g}(\sqrt{1 + (gt)^2} - 1).$$

But, to prepare the case of General Relativity, we can also 
introduce a proper time $\tau$ such that 
$$\frac{dt}{d\tau} = \frac{1}{\sqrt{1-v^2}}.$$
Like $v = dy/dt$, we can introduce the velocity $w = dy/d\tau$
and we have 
$$w = \frac{dy}{d\tau} = \frac{dy}{dt} \frac{dt}{d\tau} = 
\frac{v}{\sqrt{1-v^2}}$$
and then 
$$\frac{dw}{d\tau} =  
\frac{dw}{dv} \frac{dv}{dt} \frac{dt}{d\tau}  
= \frac{1}{\sqrt{1-v^2}^3}  g \sqrt{1-v^2}^3 \frac{1}{\sqrt{1-v^2}} 
= g \frac{1}{\sqrt{1-v^2}}$$
From $w = v/\sqrt{1-v^2}$, we get $v = w/\sqrt{1+w^2}$, thus 
\begin{align}
\frac{dw}{d\tau} =  g \sqrt{1+w^2}\label{SR1}
\end{align}
which is the equation of motion in terms of proper time. 

In the same way, we have
\begin{align}
\frac{dt}{d\tau} = \sqrt{1+w^2} \label{SR2}
\end{align}
which is the equation describing the relation between coordinate time $t$ 
and proper time $\tau$. 

Solving Equation (\ref{SR1}),  we get
$$w = \sinh(g \tau)$$
and 
$$y = \frac{1}{g} (\cosh(g \tau) - 1)$$
Equation (\ref{SR2}) then becomes 
$$\frac{dt}{d\tau} = \cosh(g\tau)$$
and integrating it, we get 
$$t = \frac{1}{g} \sinh (g\tau)$$
from which we get 
$$y = \frac{1}{g}(\sqrt{1 + (gt)^2} - 1)$$
as expected. 

Note that the velocity $w = \sinh(g \tau)$ goes to infinity when $\tau$ does. 
But the mapping from coordinate
time to proper time $\tau = (1/g) \arsinh (gt)$ 
slows down in such a way that the velocity 
$v = g t/\sqrt{1 + (gt)^2}$ remains bounded by $1$. Hence
the particle never goes faster than light.

The spacetime trajectory of the particle is a branch of the hyperbola 
of equation  
$$(gy + 1)^2 - (gt)^2 = 1$$ 
Thus, in Special Relativity,
the spacetime trajectory of a particle upon which a constant
force is applied is not a branch of a parabola, but a branch of an
hyperbola and the problem of modeling the 
motion of such a particle, with a cellular automaton, boils down to
that of the approximability of a branch of hyperbola.

The branch of hyperbola 
$$y= \frac{1}{g}(\sqrt{1 + (gt)^2} - 1)$$
has an asymptote
$$y'= t - \frac{1}{g}$$
with whom the difference is
$$y - y' =  \frac{1}{g} (\sqrt{1 + (gt)^2} - gt) =
\frac{1}{g(\sqrt{1 + (gt)^2} + gt)}$$ 
As expected, $y - y'$ decreases and goes to $0$, when $t$ goes to
infinity.  Moreover, if working with a space accuracy $\Delta$,
the hyperbola and its asymptote become indistinguishable at a time
$\theta$ verifying
$$\Delta=\frac{1}{g} \sqrt{1 + (g \theta)^2} - g \theta$$
that is at time 
$$\theta = \frac{1 - (g \Delta)^2}{2g^2\Delta}$$ 

Consider an integer $N$ and let $\Delta=(1/g)/N$.  As $N$ can be taken
as large as we wish, $\Delta$ can be taken as small as we wish.
Consider the discretization of spacetime with a temporal and spatial
step $\Delta$.  Consider the function $\tilde{y}$ from $\Delta
{\mathbb N}$ to $\Delta {\mathbb Z}$ mapping every $k \Delta$ smaller
than $\theta$ to the rounding of $y(k \Delta)$ in $\Delta {\mathbb Z}$
and every $k \Delta$ larger than $\theta$ to $y'(k \Delta) = k \Delta
- (1/g) = (k - N) \Delta$.

Let us construct a one-dimensional cellular automaton which represents
the discrete motion $\tilde{y}$.  Set the state space $\Sigma = \{q,
0, ..., L - 1, \infty\}$, with $L = \ulcorner \theta / \Delta
\urcorner$.
Let us denote by $c(k,\sigma)$ the configuration such that all cells
are in state $q$ except the cell $k$ which is in state $\sigma$. If
$\sigma \in \{0, ..., L - 1\}$, the cellular automaton maps
$c(k,\sigma)$ to either $c(k , \sigma + 1)$ or $c(k+1 , \sigma +
1)$---assuming $(L - 1) + 1 = \infty$---depending on whether
$\tilde{y}((k+1) \Delta) - \tilde{y}(k \Delta)$ is equal to zero or to
$\Delta$, and $c(k,\infty)$ to $c(k+1,\infty)$.

Note that the internal state can be seen as a clock, the state $k$
corresponding to the time $k \Delta$. It can also be seen as a
representation of the momentum, as the momentum 
$p = m' v / \sqrt{1-v^2} = m' w = 
m' g t$ grows linearly with time,
the state $k$ representing the momentum $k m' g \Delta$.  The state
$\infty$ corresponds to the case where momentum is large enough, so
that its influence on velocity can be neglected, and the motion of the
particle can be approximated by a uniform motion at the speed of light.

The number of states needed to simulate the spacetime trajectory is
$$l = 2 + \frac{\theta}{\Delta} = \frac{1}{2g^2\Delta^2} + \frac{3}{2}$$

If we assume that the number of bits that can be encoded in a cell of
length $\Delta$ is $\Delta/\rho$, for some distance $\rho$, then, to
encode $\log_2 (1/(2 g^2 \Delta^2) + 3/2)$ bits, we need 
a cell of size $\Delta$ such that
$$\log_2 (\frac{1}{2 g^2\Delta^2} + \frac{3}{2}) \leq \Delta / \rho$$
that is 
$$\Delta / \rho 
- \log_2 (\frac{1}{2 g^2\Delta^2} + \frac{3}{2}) 
\geq 0$$
The function
$\Delta / \rho 
- \log_2 (1/(2 g^2\Delta^2) + 3/2)$ 
is monotonic in $\Delta$, so this equation can be numerically solved. 

For example, if $g = 1.09~10^{-16}~\mbox{m}^{-1}$ and 
$\rho = 1.6~10^{-35}~\mbox{m}$, this equation boils down to
$$\Delta \geq 5.11~10^{-33}~\mbox{m} = 320 \rho$$ Indeed, if we take
$\Delta = 320 \rho$, a cell can encode $320$ bits and $l =
1.54~10^{96} = 2^{320}$.  

So, with an accuracy of the order of magnitude of $10^{-33}~\mbox{m}$, 
constant force in Special Relativity does not require a particle to
contain more than a few hundred bits. 

\section{Free fall in General Relativity}

In General Relativity, the gravitational effect of a body of mass $M$
at a distance $d = R - y$ is described by the metric tensor
$$\left(\begin{array}{cc}
g_{tt} & 0\\
0 & -\frac{1}{g_{tt}}
\end{array}\right)$$
where $g_{tt} = 1 - 2m / (R-y)$.

The motion of a particle is described as a function mapping its proper
time $\tau$ to a point in spacetime 
$\left(\begin{array}{c}
    t(\tau) \\
    y(\tau)
\end{array}\right)$. 
The equations of this motion are \cite{d1899introducing}:
$$\frac{d^2 t}{d\tau^2} + 2 \Gamma^t_{yt} \frac{dt}{d\tau}\frac{dy}{d\tau} =0 $$
$$\frac{d^2 y}{d\tau^2} + \Gamma^y_{tt} (\frac{dt}{d\tau})^2 + \Gamma^y_{yy} (\frac{dy}{d\tau})^2 = 0 $$
where 
$$\Gamma^y_{tt} = \frac{1}{2} g_{tt} \frac{dg_{tt}}{dy}$$
$$\Gamma^y_{yy} = - \frac{1}{2} \frac{1}{g_{tt}} \frac{dg_{tt}}{dy}$$
$$\Gamma^t_{yt} = \Gamma^t_{ty} = \frac{1}{2} \frac{1}{g_{tt}} \frac{dg_{tt}}{dy}$$
are the non-zero Christoffel symbols corresponding to this metric tensor, 
that is 
$$\frac{d^2 t}{d\tau^2} = -
\frac{1}{g_{tt}} 
\frac{dg_{tt}}{dy}
\frac{dt}{d\tau}\frac{dy}{d\tau}$$
$$\frac{d^2 y}{d\tau^2}  = - \frac{1}{2} \frac{dg_{tt}}{dy}
(g_{tt} (\frac{dt}{d\tau})^2 - \frac{1}{g_{tt}} (\frac{dy}{d\tau})^2)$$
to which we can add a third equation expressing that $\tau$ is a proper time
$$g_{tt} (\frac{dt}{d\tau})^2 - \frac{1}{g_{tt}}(\frac{dy}{d\tau})^2 = 1$$
Note that adding this third equation permits to drop the first, because 
differentiating the third equation and using the second to 
replace $d^2 y/d\tau^2$
by $- (1/2) (dg_{tt}/dy)
(g_{tt} (dt/d\tau)^2 - (1/g_{tt}) (dy/d\tau)^2)$
yields the first.
Using this third equation, the second can also be simplified to 
$$\frac{d^2y}{d\tau^2}  = - \frac{1}{2} \frac{dg_{tt}}{dy}$$
Thus, introducing the velocity $w = dy/d\tau$, the equations of motion 
boil down to the two equations
$$\frac{dw}{d\tau}  = - \frac{1}{2} \frac{dg_{tt}}{dy}$$
\begin{align}
\frac{dt}{d\tau} = \frac{1}{g_{tt}} \sqrt{g_{tt} + w^2}\label{eq:GR2gtt}
\end{align}
which are, respectively, the equation of motion in terms of proper time
and that describing the relation between coordinate time $t$ 
and proper time $\tau$. 

``Constant force due to free fall'' would make for a non-standard
concept in General Relativity. On the one hand, a constant force of
non-gravitational origin could indeed be applied in flat space and
lead to the exact same computations as Special Relativity. On the
other hand, free-falling could just mean following a geodesic
trajectory in some more or less complicated metric---although not in a
constant one. Indeed, making $g_{tt}$ constant as in the approximation
of Equation \eqref{eq:NP} into \eqref{eq:NPapprox} becomes an
over-approximation, as the geodesics then become linear. So, we define
this ``first order approximated free fall'' as the first non-trivial
approximation of the metric tensor, that is we take a linear
approximation of $g_{tt}$ as
$$g_{tt} = 1-\frac{2m}{R}-\frac{2m}{R^2} y= 1 - \frac{2m}{R} - 2gy$$
where $g = m/R^2$ is the acceleration of gravity, as before. 
Introducing $y_1 = (1 - (2m/R))/2g$ we 
get 
$$g_{tt} = 2g (y_1 - y)$$
In the same way, we approximate $dg_{tt}/dy =  - 2m/(R-y)^2$ 
by $- 2m/R^2 = -2g$. 

The equations of motion then become 
\begin{align}
\frac{dw}{d\tau}  = g\label{GR1}
\end{align}
\begin{align}
\frac{dt}{d\tau} = \frac{1}{2g(y_1-y)} \sqrt{2g(y_1 - y) + w^2}
\label{GR2}
\end{align}
Note the differences and similarities with the cases of the previous
settings.  The equation describing the relation between coordinate
time and proper time, that is Equation \eqref{GR2} or \eqref{eq:GR2gtt}
does coincide with that of Special Relativity, that is
Equation \eqref{SR2}, in the flat spacetime case when $g_{tt} = 1$. But the
equation of motion, that is Equation \eqref{GR1}, coincides not with
Special Relativity, that is Equation \eqref{SR1}, but with Newtonian
physics, that is \eqref{eq:NP1}.

Integrating Equation (\ref{GR1}), we get 
$$w =  g \tau$$
and
$$y =  \frac{1}{2} g \tau^2$$
Equation (\ref{GR2}) then becomes
$$\frac{dt}{d\tau} 
= \frac{1}{2g(y_1-(1/2)g\tau^2)} \sqrt{2g(y_1 - (1/2)g\tau^2) + (g\tau)^2}$$
$$\frac{dt}{d\tau} = \sqrt{\frac{y_1}{2g}} \frac{1}{y_1 - (1/2) g\tau^2}$$
Integrating it, we obtain 
$$t = \frac{1}{g}\artanh(\tau \sqrt{\frac{g}{2y_1}})$$
$$\tau = \sqrt{\frac{2y_1}{g}}\tanh(gt)$$
and finally
$$y=\frac{1}{2}g\tau^2=y_1(\tanh(gt))^2$$
Note that the velocity 
$$v = \frac{dy}{dt} = 2 g y_1 \tanh(gt) (1 - (\tanh(gt))^2) = 
(1 - \frac{2m}{R}) \tanh(gt) (1 - (\tanh(gt))^2)$$
is bounded by $1$, hence the particle never goes faster than light.

Like in Special Relativity, the velocity $w = g \tau$ goes to infinity 
when $\tau$ does and 
the mapping from coordinate
time to proper time 
$$\tau = \sqrt{\frac{2y_1}{g}}\tanh(gt)$$
slows down in such a way that the velocity $v$ is 
bounded by $1$. Moreover, unlike in Special Relativity,
when $t$
goes to infinity, $\tau$ has a finite limit $\sqrt{2y_1/g}$. Thus, an
infinite amount of coordinate time corresponds to a finite amount of
proper time.  As a consequence, with respect to coordinate time, after
an acceleration phase, the particle decelerates and has a limit
position $y_1$.

The distance to the limit at time $t$ is 
$$y_1 - y = y_1(1 - (\tanh(gt))^2)$$
As expected, $y_1 - y$ decreases and goes to $0$ when $t$ goes to
infinity.  Moreover, if working with a space accuracy of $\Delta$,
the position and its limit become indistinguishable at a time $\theta$
verifying
$$\Delta = y_1 (1 - (\tanh(g\theta))^2)$$
that is at time 
$$\theta = \frac{1}{g} \artanh(\sqrt{1 - \frac{\Delta}{y_1}})$$

Consider a distance $\Delta$ that can be taken as small as we
wish. Like in the case of Special Relativity, consider the
discretization of spacetime with a temporal and spatial step $\Delta$ and 
the function $\tilde{y}$ from $\Delta {\mathbb N}$ to $\Delta
{\mathbb Z}$ mapping every $k \Delta$ smaller than $\theta$ to the
rounding of $y(k \Delta)$ in $\Delta {\mathbb Z}$ and every $k \Delta$
larger than $\theta$ to $y_1$.

Let us construct a one-dimensional cellular automaton which
represents the discrete motion $\tilde{y}$.
Set the state space 
$\Sigma = \{q, 0, ..., L - 1, \infty\}$, with $L = 
\ulcorner \theta / \Delta \urcorner$. 
If $\sigma \in \{0, ..., L - 1\}$, the cellular automaton 
maps $c(k,\sigma)$ to either 
$c(k , \sigma + 1)$ or $c(k+1 , \sigma + 1)$---assuming 
$(L - 1) + 1 = \infty$---depending on whether 
$\tilde{y}((k+1) \Delta) - \tilde{y}(k \Delta)$ is equal 
to zero or to $\Delta$, and 
$c(k,\infty)$ to itself.

The number of states needed to simulate the spacetime trajectory is
$$l = 2 + \theta/\Delta
= 2 + \frac{1}{g \Delta} \artanh(\sqrt{1 - \frac{\Delta}{y_1}})$$

If we assume that the number of bits that can be encoded in a 
cell of length $\Delta$ is $\Delta/\rho$, for some distance $\rho$, 
then, to encode
this amount of information, 
we need a cell of size $\Delta$ where 
$$\log_2 (2 + \frac{1}{g \Delta} \artanh(\sqrt{1 - \frac{\Delta}{y_1}})) 
\leq \frac{\Delta}{\rho}$$
that is 
$$\frac{\Delta}{\rho} - 
\log_2 (2 + \frac{1}{g \Delta} 
\artanh(\sqrt{1 - \frac{\Delta}{y_1}})) \geq 0$$
This function is monotonic in $\Delta$, so this equation can be
numerically solved.
For example, if $m = 4.42~10^{-3}~\mbox{m}$, $R = 6.37~10^6~\mbox{m}$, 
and $\rho = 1.6~10^{-35}~\mbox{m}$, we get $g = 1.09~10^{-16}~\mbox{m}^{-1}$ 
and $y_1 = 4.57~10^{15}~\mbox{m}$. 
This equation boils down to 
$$\Delta \geq 2.69~10^{-33}~\mbox{m} = 168 \rho$$
Indeed, if we take $\Delta = 168 \rho$, a cell can encode $168$ bits
and $l = 1.92~10^{50} = 2^{168}$.

So, with an accuracy of the order of magnitude of $10^{-33}~\mbox{m}$, 
General Relativity also does not require a free falling particle to
contain more than a few hundred bits. 

\section{Conclusion}

Newtonian physics and Relativity completely differ with respect to the
possibility modelling free fall within a cellular automaton. Such a
simulation is not possible for Newtonian physics, while it is possible
both in Special---constant force---and General Relativity---geodesics
in a linearly approximated metric.  The simulation can be very
accurate with a reasonable number of internal states: a few hundred
bits suffice to achieve an accuracy of the order of magnitude of
$10^{-33}~\mbox{m}$. So, as far as free fall is concerned, Relativity
is completely consistent with the hypotheses of a bounded velocity of
propagation of information and of a bounded density of information,
unlike Newtonian physics.

In this work, we made explicit these accurate cellular automata, by
exploiting the asymptotes to the trajectory, that exist
both in Special and General Relativity. There was no need to
use auxiliary signals as in \cite{Mazoyer199953}. We have proved
the existence of such cellular automata, but made no attempt to design
``natural'' ones: the local rules use the solutions of the
equations of motion in order to know whether the particle should move,
or not. Moreover, there was clearly no attention paid to covariance. 
The design of more natural automata is of course of prime
importance. In the case of General Relativity for instance, the metric at each point ought to be carried by the corresponding cell: we began to address this question both in the classical case \cite{ArrighiGeodesics}, and, building upon \cite{di2013quantum}, in the quantum case \cite{ArrighiGRDirac}. 

\section*{Acknowledgements}

The authors thank David Janin for useful discussions on this paper,
and Alejandro P\'eres for indications on General Relativity. This work
has been funded by the ANR-12-BS02-007-01 TARMAC grant. Pablo Arrighi
is also a member of IXXI, where this research was partially conducted.

\bibliographystyle{eptcs}
\bibliography{freefall}

\end{document}